\documentclass[referee]{aa}

\usepackage{graphicx}

\newcommand{\beq}{\begin{equation}}
\newcommand{\eeq}{\end{equation}}
\newcommand{\lab}{\label}

\newcommand{\bfxi}{\mbox{\boldmath $\xi$}}

\newcommand{\bfal}{\mbox{\boldmath $\alpha$}}

\begin{document}

\title{Gravitational lensing by spherically symmetric lenses with angular momentum}

\author{M. Sereno \inst{1} \fnmsep \inst{2}, V.F. Cardone \inst{2} \fnmsep \inst{3}}

\offprints{M. Sereno, \\
\email{sereno@na.infn.it}}

\institute{Dipartimento di Scienze Fisiche, Universit\`{a} degli Studi di Napoli
``Federico II", Via Cinthia, Compl. Univ. Monte S. Angelo, 80126
Napoli, Italia
\and Istituto Nazionale di Fisica Nucleare, Sez. Napoli, Via Cinthia, Compl.
      Univ. Monte S. Angelo, 80126 Napoli, Italia
\and Dipartimento di Fisica ``E.R. Caianiello", Universit\`{a} di Salerno, Via S. Allende, 84081
Baronissi, Salerno, Italia}

\date{Received / Accepted}

\titlerunning{Gravitational lensing and gravitomagnetism}
\authorrunning{M. Sereno, V.F. Cardone}

\abstract{Following Sereno (\cite{io02}), we discuss the bending of light
rays by spherically symmetric lenses with angular momentum. For
several astrophysical systems, such as white dwarfs and galaxies,
gravitomagnetism induces a correction on the deflection angle as large
as $0.1\%$.
}

\maketitle

\section{Introduction}

Gravitational lensing is one of the best investigated phenomena of
gravitation. In the framework of general relativity, its lowest-order
predictions have been confirmed by observative astrophysics on very
different scales. On the other hand, the impressive development of
technical capabilities demands for a full treatment of lensing theory
to any order of approximation. The study of higher order perturbative
terms is the link between weak and strong regimes of the theory.

Mass-energy currents relative to other masses generate space-time
curvature. This phenomenon, known as intrinsic gravitomagnetism, is a
new feature of general relativity and other conceivable alternative
metric theories of gravity and cannot be deduced by a motion on a
static background (for a detailed discussion on gravitomagnetism we
refer to Ciufolini \& Wheeler (\cite{ci+wh95})). In particular, the
effect of the angular momentum of the deflector has been studied by
several authors (Epstein \& Shapiro \cite{ep+sh80}; Ib\'{a}\~{n}ez \& Mart\'{\i}n
\cite{ib+ma82}; Ib\'{a}\~{n}ez \cite{iba83}; Dymnikova \cite{dym86};
Glicestein \cite{gli99}; Sereno \cite{io02}).

One of us (Sereno \cite{io02}) showed as the gravitomagnetic
correction to the lensing quantities can be evaluated in the usual
framework of lensing theory (see also Capozziello et al.
\cite{cap+al99}), i.e. {\it i)} weak field and slow motion
approximation for the lens; {\it ii)} thin lens hypothesis (Schneider
et al. \cite{sef}; Petters et al. \cite{pet+al01}).

In this letter, we consider the gravitomagnetic contribution to the
deflection angle for extended gravitational lenses of astrophysical
interest. We discuss spherically symmetric deflectors.

\section{The deflection angle}

As derived in Sereno (\cite{io02}), the deflection angle, to the order
$c^{-3}$, is
\beq
\lab{wf18}
\bfal(\bfxi ) \equiv \frac{4 G}{c^2}\int_{\Re^2}d^2\xi^{'}\Sigma(\bfxi^{'})
\left( 1-2\frac{\langle {\bf v}{\cdot}{\bf e}_{in}\rangle_l(\bfxi^{'})}{c}\right)
\frac{\bfxi -\bfxi^{'}}{|\bfxi -\bfxi^{'}|^2};
\eeq
$\bfxi$ is the bi-dimensional position vector in the lens plane,
orthogonal to the line of sight $l$; ${\bf e}_{in}$ is the incoming
light ray direction; $G$ is the Newton's constant of gravitation; $c$
is the speed of light. $\Sigma$ is the surface mass density  projected
along the line of sight
\beq
\lab{wf11}
\Sigma(\bfxi)\equiv \int \rho(\bfxi,l)\ dl;
\eeq
$\langle {\bf v}{\cdot}{\bf e}_{in}\rangle_l$ is the weighted average, along
the line of sight, of the component of the velocity $\bf v$ of the
mass element orthogonal to the lens plane,
\beq
\lab{wf12}
\langle {\bf v}{\cdot}{\bf e}_{in}\rangle_l (\bfxi)\equiv \frac{\int ({\bf v}(\bfxi,l){\cdot} {\bf e}_{in}) \
\rho(\bfxi,l)\ dl}{\Sigma(\bfxi)}.
\eeq

Let us consider a spherically symmetric lens that rotates about an
arbitrary axis, $\hat{\eta}$, passing through its center (i.e. a main
axis of inertia). To specify the orientation of the rotation axis, we
need two Euler's angles: $\alpha$ is the angle between $\hat{\eta}$
and the $\xi_2$-axis; $\beta$ is the angle between the line of sight
$\hat{l}$ and the line of nodes defined at the intersection of the
$\hat{l\ \xi_1}$ plane and the equatorial plane (i.e., the plane
orthogonal to the rotation axis and containing the lens center). Using
the axial symmetry about the rotation axis, we find
\beq
{\bf v}{\cdot}{\bf e}_{in}(\xi_1,\xi_2,l)=
-\omega (R) \left[ \xi_1 \cos \alpha+\xi_2 \sin \alpha \cos \beta \right]
\equiv -\omega_2 (R) \xi_1 + \omega_1 (R)\xi_2,
\eeq
where $\omega (R)$ is the modulus of the angular velocity at a
distance $R \equiv (R_1^2+R_2^2)^{1/2}$ from the rotation axis;
$\hat{R}_1$ (that, given the spherical symmetry of the system, can be
taken along the line of nodes) and $\hat{R}_2$ are the axes on the
equatorial plane; $\omega_1$ and $\omega_2$ are the components of
$\omega$ along, respectively, the $\xi_1$- and the $\xi_2$-axes. It is
\beq
R_1=l\cos \beta+\xi_1 \sin \beta,
\eeq

\beq
R_2=-l \cos \alpha \sin \beta + \xi_1 \cos \alpha \cos \beta +\xi_2
\sin \alpha .
\eeq
Let us assume a rigid rotation, $\omega (R) =\omega = const$. We have
\beq
\langle {\bf v}{\cdot}{\bf e}_{in}\rangle_l = -\omega_2  \xi_1 + \omega_1 \xi_2.
\eeq
We can, now, evaluate the integral in Eq.(\ref{wf18}); it is,
\beq
\lab{circ1}
\alpha_1(\xi, \theta)=\frac{4G}{c^2}\left\{ \frac{M(\xi)}{\xi}\cos \theta+
\frac{I_N(\xi)}{\xi^2}\left( \frac{\omega_2}{c} \cos 2\theta -\frac{\omega_1}{c} \sin 2\theta \right)
-M(>\xi)\frac{\omega_2}{c} \right\};
\eeq
\beq
\lab{circ2}
\alpha_2(\xi, \theta)=\frac{4G}{c^2}\left\{ \frac{M(\xi)}{\xi}\sin \theta +\frac{I_N(\xi)}{\xi^2}
\left( \frac{\omega_1}{c} \cos 2\theta +\frac{\omega_2}{c} \sin 2\theta \right)+ M(>\xi)\frac{\omega_1}{c} \right\}.
\eeq
$\xi$ and $\theta$ are polar coordinates in the lens plane; $M(\xi)$
is the mass of the lens within $\xi$,
\beq
M(\xi) =2 \pi \int_0^\xi \Sigma(\xi^{'}) \xi^{'} d \xi^{'};
\eeq
$M(>\xi)$ is the lens mass outside $\xi$, $M(>\xi)
\equiv M(\infty)-M(\xi)$; and
\beq
I_N(\xi)=2 \pi \int_0^\xi \Sigma(\xi^{'}){\xi^{'}}^3 d \xi^{'}
\eeq
is the momentum of inertia of the mass within $\xi$ about a central
axis. $I_N {\times} \omega_i$ is the component of the angular momentum along
the $\xi_i$-axis.

The gravitomagnetic correction consists of the last two terms in
Eqs.(\ref{circ1},\ref{circ2}), both proportional to some components of
the angular velocity. Spherical symmetry is broken. In the first
contribution, the angular momentum appears; the second one is
proportional to the mass outside $\xi$ and can be significant for
lenses with slowly decreasing mass density.

\section{The homogeneous sphere}
The gravitational phenomena connected to intrinsic gravitomagnetism
are generated by mass-energy currents relative to other masses. The
simplest lens model, the point-like Schwarzschild lens, cannot produce
such a peculiar effect since the local Lorentz invariance on a static
background does not account for the dragging of inertial frames
(Ciufolini \& Wheleer \cite{ci+wh95}). General relativity is a
classical-nonquantized theory where the classical angular momentum of
a particle goes to zero as its size goes to zero. To consider the
gravitomagnetic field, we need a further step after the point mass as
a lens model, the homogeneous sphere. Let us consider a homogeneous
sphere of radius $R$ and volume density $\rho_0$. It is
\beq
\Sigma (\xi)= 2 \rho_0 \sqrt{R^2-\xi^2},\ {\rm if}\ \xi \leq R,
\eeq
or $\Sigma (\xi) = 0$ elsewhere;
\beq
M(\xi)=M_{TOT}\left[ 1-\left(1-
\left(\frac{\xi}{R}\right)^2\right)^{3/2}\right],\ {\rm if}\ \xi \leq
R,
\eeq
or $M(\xi)=M_{TOT}$ elsewhere, $M_{TOT}=\frac{4}{3}\pi R^3 \rho_0 $;
\beq
I_N(\xi)=I_N^{TOT}\left[ 1- \left(1-
\left(\frac{\xi}{R}\right)^2\right)^{1/2} \left(
1+\frac{1}{2}\left(\frac{\xi}{R}\right)^2
-\frac{3}{2}\left(\frac{\xi}{R}\right)^4  \right) \right],\ {\rm if}\
\xi \leq R,
\eeq
or $I_N(\xi) = I_N^{TOT}$ elsewhere, $I_N^{TOT}=\frac{8}{15}\pi R^5
\rho_0$;

For light rays outside the lens, $\xi > R$, the deflection angle
reduces to
\beq
\lab{hom1}
\alpha_1(\xi, \theta)=\frac{4G}{c^2}\left\{ \frac{M_{TOT}}{\xi}\cos \theta +\frac{ I_N^{TOT}}{\xi^2}
\left( \frac{\omega_2}{c} \cos 2\theta -\frac{\omega_1}{c} \sin 2\theta \right) \right\},
\eeq

\beq
\lab{hom2}
\alpha_2(\xi, \theta)=\frac{4G}{c^2}\left\{ \frac{M_{TOT}}{\xi}\sin \theta +
\frac{I_N^{TOT}}{\xi^2} \left( \frac{\omega_1}{c} \cos 2\theta +\frac{\omega_2}{c} \sin 2\theta \right)  \right\}.
\eeq

The gravitomagnetic correction is significant if
\beq
\frac{I_N^{TOT}}{M_{TOT}}\frac{\omega }{c \ \xi}=\frac{J}{ M_{TOT} \ c \ \xi} \stackrel{>}{\sim } 10^{-3},
\eeq
where $J \equiv I_N {\times} \omega$ is the angular momentum. To have a
non-negligible gravitomagnetic effect, the angular momentum of the
lens has to be non-negligible compared to the angular momentum of a
particle of mass $M_{TOT}$ and velocity $c$ in a circular orbit of
radius $\xi$ around the rotation axis.

Let us consider a lens rotating about the $\xi_2$-axis  ($\omega_1
=0$, $\omega_2 =\omega$) and a light ray in the equatorial plane, $\theta =0$. The
deflection generated by the gravitomagnetic field is
\beq
\alpha_{GRM} = \frac{4G}{c^3}\frac{J}{\xi^2}.
\eeq
The sun bends a light ray grazing its limb, $\xi =R_{\odot}$, by
$1.75$ arcsec. Given its angular momentum, $J_{\odot}\simeq 1.6{\times}
10^{48}$g cm$^2$s$^{-1}$ (Allen \cite{all83}), the gravitomagnetic
correction is $\sim 0.7$ $\mu$arcsec (see also Epstein \& Shapiro
(\cite{ep+sh80})).

For an early type star,
$J=10^2J_{\odot}\left(\frac{M}{M_{\odot}}\right)^{5/3}$ (Kraft
\cite{kra67}). For $M=1.4 M_{\odot}$, $R=1.1R_{\odot}$ and for a light
ray grazing the limb, $\alpha_{GRM}
\simeq 0.1$ milliarcsec, that is, a correction of $\sim 4 {\times} 10^{-5}$.

The gravitomagnetic field becomes more significant for a fast rotating
white dwarf, where $J \sim \sqrt{0.2 G M^3 R}$ (Padmanabhan
\cite{pad99}).For $M
\sim M_\odot$, $R \sim 10^{-2}R_\odot$, $\xi \sim 6 R$, $\alpha_{GRM}
\simeq 0.03$ arcsec, that is, a correction of $\sim 10^{-3}$.

\section{The isothermal sphere}
Isothermal spheres (ISs) are widely used in astrophysics to model
systems on very different scales, from galaxy haloes to clusters of
galaxies; also, IS can be adopted to study microlensing by non-compact
invisible objects in the halo (Sazhin et al. \cite{saz+al96}).

Let us consider an IS with a finite core radius $\xi_c$. The surface
density is
\beq
\Sigma^{IS} (\xi)=\frac{\sigma^2_v}{2G}\frac{1}{(\xi^2+\xi_c^2)^{1/2}},
\eeq
where $\sigma_v$ is the velocity dispersion. We have,
\beq
M^{IS}(\xi)=\frac{\pi \sigma^2_v}{G}\xi_c \left[ \left( 1+
\left( \frac{\xi}{\xi_c} \right)^2 \right)^{1/2} -1
    \right],
\eeq

\beq
I_N^{IS}(\xi) = \frac{\pi \sigma^2_v}{3G}\xi_c^3 \left[ 2+  \left(
\left( \frac{\xi}{\xi_c} \right)^2 -2 \right) \left( 1+
\left( \frac{\xi}{\xi_c} \right)^2 \right)^{1/2}
\right];
\eeq
When $\xi_c=0$, we have the singular isothermal sphere (SIS). Then,
\beq
M^{SIS}(\xi)=\frac{\pi \sigma^2_v}{G}\xi ,
\eeq

\beq
I_N^{SIS}(\xi) = \frac{\pi \sigma^2_v}{3G}\xi^3 .
\eeq
Since the total mass is divergent, we introduce a cut-off radius $R \gg
\xi$. For the SIS, the deflection angle reduces to
\beq
\alpha_1^{SIS}(\xi, \theta)=4 \pi \left( \frac{\sigma_v}{c} \right)^2 \left\{   \cos \theta  +
\frac{\omega_2}{c}\left[ \xi \left( \frac{\cos 2 \theta}{3} +1 \right)  -R  \right]
-\frac{\omega_1}{c}\xi  \frac{\sin 2 \theta}{3}   \right\},
\eeq

\beq
\alpha_2^{SIS}(\xi, \theta)=4 \pi \left( \frac{\sigma_v}{c} \right)^2 \left\{   \sin \theta  +
\frac{\omega_1}{c}\left[ \xi \left( \frac{\cos 2 \theta}{3} -1 \right)  +R  \right]
+\frac{\omega_2}{c}\xi  \frac{\sin 2 \theta}{3}   \right\}.
\eeq
The correction couples kinematics, through the angular velocity, and
geometry, through the cut-off radius. As can be easily seen, the gravitomagnetic
effect is significant when
\beq
\frac{\omega}{c}R \stackrel{>}{\sim}10^{-3};
\eeq
In particular, in the inner regions ($\xi \ll R$), the above equations
reduce to
\beq
\alpha_1^{SIS}(\xi \ll R, \theta)=4 \pi \left( \frac{\sigma_v}{c} \right)^2 \left\{ \cos \theta
-\frac{R \omega_2}{c}   \right\},
\eeq

\beq
\alpha_2^{SIS}(\xi \ll R, \theta)=4 \pi \left( \frac{\sigma_v}{c} \right)^2 \left\{ \sin \theta
+\frac{R \omega_1}{c}   \right\};
\eeq
the correction derives from the mass outside the considered radius.

We can model a typical galaxy as a SIS with $\sigma_v \sim 200$ km
s$^{-1}$, $R \stackrel{<}{\sim} 50$ kpc and $J \equiv I_N(R){\times} \omega
\sim 0.1 M_{\odot}$ kpc$^2$s$^{-1}$, as derived from numerical
simulations (Vitviska et al. \cite{vit+al01}). It is,
\beq
\frac{\omega}{c}R \sim \frac{G}{c^3} J \left( \frac{c}{\sigma_v} \right)^2 R^{-2} \sim 10^{-3}.
\eeq
The gravitomagnetic correction is quite significant, increases with
the ordered motion of the stars (i.e., with the angular momentum) and
decreases with the random proper motions (i.e., with the dispersion
velocity).

\section{Power law models}

Power law models can be considered as a generalization of the IS
(Schneider et al. \cite{sef}) and are often adopted to model mass
distribution in clusters of galaxies by lensing inversion (Sereno
\cite{io02cl}). It is
\beq
\Sigma^{PL}= \Sigma_0 \frac{1+p \left( \frac{\xi}{\xi_c} \right)^2}{\left(
1+\left( \frac{\xi}{\xi_c} \right)^2\right)^{2-p}};
\eeq
the slope parameter $p$ determines the softness of the mass profile of
the lens. A power law model with $p=1/2$ approximates the isothermal
sphere at large radius. It is,
\beq
M^{PL}(\xi)= \pi \Sigma_0 \xi^2 \left[ 1+\left( \frac{\xi}{\xi_c}
\right)^2\right]^{p-1},
\eeq
\beq
I_N^{PL}(\xi)= \frac{ \pi \Sigma_0 \xi_c^4}{p(1+p)}
\left[ \left( 1+\left( \frac{\xi}{\xi_c}
\right)^2\right)^{p-1} \left(1+(1-p)\left( \frac{\xi}{\xi_c}
\right)^2 +p^2\left( \frac{\xi}{\xi_c}
\right)^4   \right) -1 \right].
\eeq

\section{Summary and discussion}
We have investigated the effect of dragging of inertial frames in
gravitational lensing for spherically symmetric lenses and a general
expression for the deflection angle, to the order $c^{-3}$, has been
derived. We have explicitly considered isothermal spheres, power law
models and the homogeneous sphere. Both for galaxies and white dwarfs,
the gravitomagnetic correction can be as large as $0.1\%$.

The satellite Hipparcos, launched in 1989 by ESA, can measure the
position of stars with accuracy of nearly a milliarcsec. New
generation space interferometric mission, such as SIM by NASA
(scheduled for launch in 2009), should greatly improve this accuracy.
Measurements of deflection of electromagnetic waves could give one of
the first experimental evidences of gravitomagnetism.


\begin{thebibliography}{99}

\bibitem[1983]{all83}
Allen C.W., Astrophysical Quantities, 1983, The Athlone Press, London

\bibitem[1999]{cap+al99}
Capozziello, S., Lambiase, G. Papini, G. Scarpetta, G., 1999, Phys.
Lett. A 254, 11.

\bibitem[1995]{ci+wh95}
Ciufolini I., Wheeler J.A., 1995, Gravitation and Inertia,Princeton
University Press, Princeton

\bibitem[1986]{dym86}
Dymnikova I., 1986, Relativity in Celestial Mechanics and Astrometry,
eds. Kovalevsky J., Brumberg A., 411.

\bibitem[1980]{ep+sh80}
Epstein R., Shapiro I., 1980, Phys. Rev. D 22, 2947.

\bibitem[1999]{gli99}
Glicenstein J.F., 1999, A\&A 343, 1025.

\bibitem[1983]{iba83}
Ib\'{a}\~{n}ez J., 1983, A\&A 124, 175.

\bibitem[1982]{ib+ma82}
Ib\'{a}\~{n}ez J., Mart\'{\i}n J., 1982, Phys. Rev. D 26, 384.

\bibitem[1967]{kra67}
Kraft R.P., 1967, ApJ 150, 551

\bibitem[2001]{pad99}
Padmanabhan T., 2001, Theoretical Astrophysics Vol. II, Cambridge
University Press, Cambridge

\bibitem[2001]{pet+al01}
Petters A.O., Levine H., Wambsganss J., 2001, Singularity Theory and
Gravitational Lensing, Birkh\"{a}user, Boston

\bibitem[1996]{saz+al96}
Sazhin M.V., Yagola A.G., Yakubov A.V., 1995, Phys. Lett. A 208, 276

\bibitem[1992]{sef}
Schneider P., J. Ehlers J., Falco E.E., 1992, Gravitational Lenses,
(Springer-Verlag, Berlin)

\bibitem[2002a]{io02}
Sereno M., 2002a, [astro-ph/0209148]

\bibitem[2002b]{io02cl}
Sereno M., 2002b, A\&A in press; [astro-ph/0209210]

\bibitem[2001]{vit+al01}
Vitviska M., Klypin A.,Kravtsov A.V., et al., 2001, ApJ, submitted;
astro-ph/0105349

\end{thebibliography}
\end{document}